\documentclass[%
 reprint,
 amsmath,amssymb,
 aps,prl,
]{revtex4-2}

\usepackage{graphicx}
\usepackage{dcolumn}
\usepackage{bm}

\usepackage[dvipsnames]{xcolor}

\begin{document}

\preprint{APS/123-QED}

\title{Optimal Moment-based Characterization of a Gaussian State}

\author{Niels Tripier-Mondancin}
\author{Ilya Karuseichyk}
\author{Mattia Walschaers}
\author{Valentina Parigi}
\author{Nicolas Treps}
\affiliation{%
 Laboratoire Kastler Brossel, Sorbonne Université, CNRS, ENS-Université PSL, Collège de France, 4 place Jussieu, Paris F-75252, France
}%

\date{\today}

\begin{abstract}
Fast and precise characterization of Gaussian states is crucial for their effective use in quantum technologies. In this work, we apply a multi-parameter moment-based estimation method that enables rapid and accurate determination of squeezing, antisqueezing, and the squeezing angle of the squeezed vacuum state. Compared to conventional approaches, our method achieves faster parameter estimation with reduced uncertainty, reaching the Cramér-Rao bound. We validate its effectiveness using the two most common measurement schemes in continuous-variable quantum optics: homodyne detection and double homodyne detection. This rapid estimation framework is well-suited for dynamically characterizing sources with time-dependent parameters, potentially enabling real-time feedback stabilization.
\end{abstract}

\maketitle


Squeezed states have been utilized in quantum optics laboratories for more than 35 years, and are at the heart of the development in quantum optics. They found their application in various branches of quantum metrology \cite{dodonov_nonclassical_2002, schnabel_squeezed_2017,lawrie_quantum_2019} ranging from phase estimation \cite{bondurant_squeezed_1984,combes_states_2004,pezze_quantum_2014,yu_quantum_2020}, including gravitational wave detection \cite{grote_first_2013,barsotti_squeezed_2018,tse_quantum-enhanced_2019,virgo_collaboration_increasing_2019
}, to spatial \cite{kolobov_sub-shot-noise_1993,soh_label-free_2023} and temporal \cite{patera_quantum_2019} imaging, displacement measurement \cite{treps_quantum_2003,treps_nano-displacement_2004}, clock synchronization \cite{giovannetti_quantum-enhanced_2001}, etc. They are also known to be valuable for quantum communication \cite{gottesman_secure_2003,usenko_unidimensional_2018,laudenbach_continuous-variable_2018,pirandola_advances_2020}, quantum teleportation \cite{furusawa_unconditional_1998,bowen_experimental_2003} and other quantum information protocols \cite{braunstein_quantum_2005,yonezawa_continuous-variable_2010,weedbrook_gaussian_2012}.

Homodyne detection has been a pivotal resource in characterizing these state in quantum optics. Standard methods of measuring the amount of squeezing involve sweeping the phase of the local oscillator (LO) and only looking at the highest and lowest variance of the quadrature, fitting the variance curve to extract the parameters, or using maximum likelihood estimation (MLE) to fully reconstruct the state \cite{RevModPhys.81.299}. Finally one can lock the phase on the squeezed and anti-squeezed quadrature respectively, which leads to the most accurate estimation. However these methods are either slow, difficult to implement experimentally, or do not take full advantage of the information present in the measurement. 

Another approach recently developed is to use machine learning to estimate directly the parameters of a squeezed state after feeding noisy data of a quadrature sequence to a reconstruction model based on machine learning \cite{hsieh_extract_2022,hsieh_direct_2022}. This allows for a fast and efficient estimation of the parameters, enabling feedback control. The primary limitation is the inability to evaluate the accuracy of the results, as neural networks generate estimators without error bars.

In this work, we construct and experimentally test a computationally feasible data-processing method that fully exploits the measured data and estimates multiple parameters. 
Our estimator is based on the first non-trivial statistical moments of the measured observables, which makes it fast and easy to compute. We experimentally demonstrate that the precision of our estimator saturates the  Cramér Rao Bound (CRB). The developed approach has the potential to dynamically characterize sources with varying parameters, enabling feedback control for the stabilization of such sources.\\

\begin{figure}
    \centering
    \includegraphics[width=1.0\linewidth]{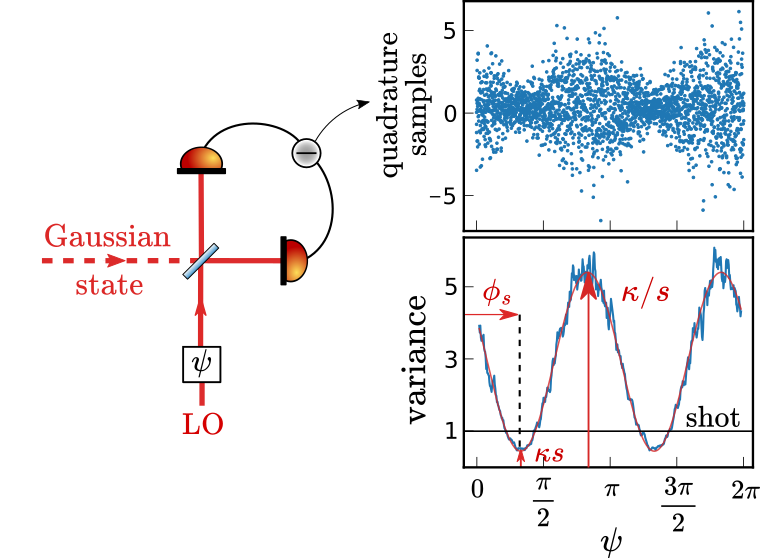}
    \caption{Homodyne detection scheme for measuring field quadratures. Top-right plot: collected quadrature samples for various phase $\psi$ of the LO. Bottom-right: sample variance of the subsets of collected data.}
    \label{fig:HD_scheme}
\end{figure}

\paragraph{Gaussian states}--- Gaussian quantum states are characterized by Gaussian statistics of the field quadratures $\hat q_\psi = \hat a_m e^{-i \psi} + \hat a_m^\dagger e^{i \psi}$. The variance of the quadrature $\hat q_\psi$ in this case can be parameterized by the three parameters $\vec \theta = (s,\kappa,\phi_s)^T$ as follows:
\begin{equation} \label{eqn:V} \Delta^2 q_\psi =V(\psi,\vec \theta)=\kappa s \cos^2 [\psi-\phi_s] + \dfrac{\kappa}{s} \sin^2[\psi-\phi_s], \end{equation}
where $\kappa \geq 1$ defines the thermal contribution to the state (the state's purity is $P=1/\kappa$), the phase $\phi_s$ defines the squeezed quadrature (the quadrature $\hat q_{\phi_s}$ has the lowest variance, equal to $\kappa s$), and $s$ defines the ratio of variances for squeezed and anti-squeezed quadratures ($0 < s \leq 1$), as shown in Fig. \ref{fig:HD_scheme}. The level of sub-shot-noise squeezing of a given Gaussian state is typically expressed in decibels as $ L_s = 10 \log_{10}[\kappa s]$. In this study, we focus on single-mode Gaussian states with zero mean-field ($\langle \hat q_\psi \rangle = 0$). Such a state is fully characterized by the three parameters $\vec \theta = (s,\kappa,\phi_s)^T$.

\paragraph{Quadrature detection}---
The characterization of a Gaussian state can be performed based on the measurement of its quadratures using homodyne detection (Fig. \ref{fig:HD_scheme}). By varying the phase $\psi$ of the local oscillator (LO), one can select the quadrature $\hat q_\psi$ to be measured. The simplest approach involves identifying the quadratures with the lowest and highest variances, and then determining the parameters $\vec \theta$ based on these variances. However, this straightforward approach underutilizes the full information available from the measurement data, and in the following sections we analyze more advanced estimators for the parameters $\vec \theta$.

\paragraph{Fisher information}---
First, let us find the total information about the parameters $\vec \theta$ available in the homodyne data. To do so, we consider a set quadrature measurements $\{\hat q_{\psi_j} \},~ j=1,...N_\psi$ performed over independent samples of the studied state. Each of them follows Gaussian statistics $\mathcal{N} (0, V(\psi_j,\vec \theta))$, thus the Fisher information (FI) matrix can be calculated as \cite{kay_fundamentals_1998}:
\begin{equation}    
    \label{eqn:FI}
        \mathcal{F}_{\alpha \beta} =\sum_{j=1}^{N_\psi} \frac{1}{2 V^2(\psi_j,\vec \theta)} \frac{\partial V(\psi_j,\vec \theta)}{\partial \theta_\alpha} \frac{\partial V(\psi_j,\vec \theta)}{\partial \theta_\beta}
\end{equation}
for indices $\alpha,\ \beta=1,2,3$. For large enough $N_\psi$, we can accurately approximate this sum by an integral. We assume that the quadratures are chosen uniformly in the range $\psi \in [0, \pi]$. 

For any estimator $\tilde \theta_\alpha$ of the parameter $\theta_\alpha$, the Cramér-Rao bound (CRB) constraints its variance such that $\Delta^2 \tilde \theta_\alpha  \ge [\boldsymbol{\mathcal{F}}^{-1}]_{\alpha \alpha}$. The explicit computation of this bound for the homodyne detection gives:
\begin{equation} \label{eqn:CRB}
    \Delta^2 \vec{\tilde \theta}  \ge \frac{1}{ N_{\psi}} \begin{pmatrix}
            s (1+s)^2 \\[6pt] 
            ~\kappa^2 \dfrac{1+s^2}{s} \\[6pt] 
            \dfrac{s}{(1-s)^2}
          \end{pmatrix},
\end{equation}
It can be used as a benchmark to evaluate the efficiency of different approaches for the estimation of the parameters $\vec \theta$.\\

\paragraph{Least squares fit estimator}---
A simple intuitive estimator of the parameters $\vec \theta$ can be obtained by dividing the quadrature samples into subsets, and then calculating the variance of each subsets to obtain the bottom plot of Fig. \ref{fig:HD_scheme}. The estimator is obtained by fitting the curve with the function $V(\psi,\vec \theta)$. 
This fit can be performed with the Least Squares (LS) method by minimizing the square difference between the fitting curve and the experimental data points:
\begin{equation} \label{eqn:LS}
    \vec{\tilde \theta}^{F} = \operatorname{arg}~ \min_{\vec \theta} \sum_{j=1}^{N_\psi} (q_j^2-V(\psi_j,\vec \theta))^2,
\end{equation}
where $q_j$ represents the measurement result of the quadrature $\hat q_{\psi_j}$. Here each subset contains only one data-point, and absence of the mean-field is taken into account.  
 
Due to the trigonometric structure of the fitting function $V(\psi,\vec \theta)$ \eqref{eqn:V}, the Fit estimator can be constructed analytically using the Fourier analysis (see appendix). This allows us to find the variance of this estimator using the error propagation principle:
\begin{equation} \label{eqn:F_var}
    \Delta^{2} \tilde \theta^F_\alpha  = \frac{1}{ N_{\psi}} \begin{pmatrix}
            \dfrac{1 + 6s^{2} + 18s^{4} + 6s^{6} + s^{8}}{8s^{2}} \\[6pt] 
            ~\kappa^2 \dfrac{1 - 2s^{2} + 18s^{4} - 2s^{6} + s^{8}}{8s^{4}}  \\[6pt] 
             \dfrac{5+6s^2 + 5s^4}{4(1-s^2)^2}
          \end{pmatrix} .    
\end{equation}
A straightforward analysis of this expression shows that this estimator never saturates the CRB \eqref{eqn:CRB}. Thus, this intuitive and simple Fit-based estimator does not extract all the available information from the homodyne data.\\ 

\paragraph{Moment-based estimator}---
The method of moments constructs the estimators for the parameters by equating the measured moments of the observables with their theoretical models \cite{kay_fundamentals_1998,gessner_multiparameter_2020}. 
Since the first moment of the homodyne data is trivial, we use the second moment to construct the estimator. Since there are more measured quadratures $\hat q_{\psi_j}$ than the parameters $\theta_\alpha$, the estimator is built using a linear combination of different observables with coefficients $c_\alpha(\psi_j, \vec \theta_0)$. Hence the Moment-based estimators $\vec{\tilde{ \theta}}^{MoM}$ are obtained as a solution of the system
\begin{equation}\label{eq:MoM estimator}
    \sum_{j=1}^{N_\psi} c_\alpha (\psi_j, \vec \theta_0)~q_j^{2} =  \sum_{j=1}^{N_\psi} c_\alpha (\psi_j, \vec \theta_0)~V(\psi_j, \vec{\tilde{ \theta}}^{MoM}),
\end{equation}
with $q_j$ the obtained quadrature samples, whereas $V$ are the theoretical moments defined in eq. \eqref{eqn:V}. 
The variance of these estimators depends on the choice of the  weights $ c_\alpha(\psi_j, \vec \theta_0)$. The method of moments gives the optimal weights \cite{gessner_multiparameter_2020}, in our case we find:
\begin{equation}
\label{eqn:c_formula}
    c_\alpha(\psi_j, \vec \theta_0)= \left(\frac{1}{2 V^2(\psi_j,\vec \theta)} \frac{\partial V(\psi_j,\vec \theta)}{\partial \theta_\alpha} \right) \Big|_{\vec \theta = \vec \theta_0}.
\end{equation}
This expression depends on a prior $\vec \theta_0$ on the parameters, and is optimal only if the prior is close to the true value of the parameters $\vec \theta$. Often, in practical situations, no a priori information is available on the true value of the parameters. In this case one can apply this scheme iteratively, starting from an arbitrary prior, and it usually converges within a few iterations. 

The analytical solution of the system of equations (\ref{eq:MoM estimator}), i.e. the explicit formula for the estimator $\vec{\tilde \theta}$, is provided in the appendix. There, we also show that the variance of this estimator is defined by the moment matrix, which coincides with the FI matrix in this case, i.e.,
\begin{equation}
    \Delta^2 \tilde \theta_\alpha^{MoM}  =[\boldsymbol{\mathcal{F}}^{-1}]_{\alpha \alpha}. 
\end{equation}
Thus the moment-based estimator saturates CRB, extracting all the available information from the homodyne data.\\

\begin{figure}
    \centering
    \includegraphics[width=0.6\linewidth]{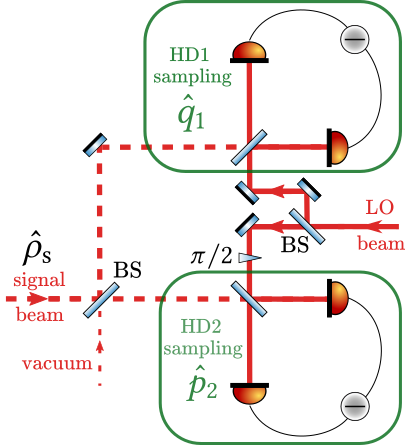}
    \caption{Double homodyne detection enables the simultaneous measurement of both quadratures but introduces additional vacuum noise.}
    \label{fig:DHD}
\end{figure}

\paragraph{Double homodyne detection}---
Another measurement technique that can be used to characterize a Gaussian state is double homodyne detection (DHD). In this case, the studied state is split in two with a beamsplitter, and two orthogonal quadratures are measured at the outputs (see Fig. \ref{fig:DHD}). Since an additional vacuum mode enters the system, the covariance matrix of the measured quadratures is
 $\mathbf{\Gamma} ({ \hat {q}}_{1}, { \hat {p}}_{2}) = \mathbf{\Gamma}_{\vec\theta} + \mathbf{I}$, with $\mathbf{\Gamma}_{\vec\theta}$ the quadrature covariance matrix of the initial state, that depends on the true values of the parameters. DHD also has Gaussian statistics, thus the FI matrix can be found as \cite{kay_fundamentals_1998}
\begin{equation}
     \mathcal{F}_{\alpha \beta} = \dfrac{1}{2} \text{Tr} \left[\mathbf{\Gamma}^{-1} \dfrac{\partial\mathbf{\Gamma}}{\partial \theta_{\alpha}}\mathbf{\Gamma}^{-1}\dfrac{\partial\mathbf{\Gamma}}{\partial \theta_{\beta}}\right].
\end{equation}
The explicit calculation of the FI provides the following bound on the variance of the DHD-based estimators:
\begin{equation} \label{eqn:DHD bound}
    \Delta^2 \vec{\tilde \theta}  \ge \frac{1}{ N_{\psi}} \begin{pmatrix}
             \dfrac{s^4+2 \kappa  s^3+2 \kappa ^2 s^2+2 \kappa  s+1}{2 \kappa ^2} \\[6pt] 
            \kappa ^2+\dfrac{s^2}{2}+\dfrac{1}{2 s^2}+\kappa  s+\dfrac{\kappa }{s} \\[6pt] 
            \dfrac{s (\kappa +s) (1+\kappa  s)}{\kappa ^2 \left(1-s^2\right)^2}
          \end{pmatrix}.
\end{equation}

Using a double homodyne detection scheme there is no need to scan the phase since two orthogonal quadratures are directly measured, instead, one repeats the measurement $\mu$ times to accumulate the statistics. The parameters of the squeezed state, including the squeezing direction, should remain constant during data acquisition. Therefore, in the absence of phase locking, measurements must be performed within a short time.  Calculating the statistical moments $\langle \hat q_1^2\rangle,~\langle \hat p_2^2\rangle,$ and $\langle \hat q_1 \hat p_2\rangle$ of the measured quadratures, one obtains directly an estimator of the covariance matrix $\mathbf{\Gamma}$. The estimator for the parameters $\vec \theta$ can be obtained from the eigensystem of the quadrature covariance matrix $\mathbf{\Gamma}_{\vec\theta} = \mathbf{\Gamma} - \mathbf{I}$. The variance of this estimator saturates corresponding CRB~\eqref{eqn:DHD bound}.

\paragraph{Experimental results}---
In our experiment, the squeezed light is created with an optical parametric oscillator (OPO). The reference value of the squeezing produced is measured by locking the phase between the squeezing and the LO and measuring the variance of the signal. The highest squeezing measured in our experiment is $L_{ref} = 3.4$ dB (See methods for details). 

The homodyne data is acquired with a phase $\psi_{j}$ uniformly distributed within the $n\pi$ range, $n \in \mathbb{N}$. We choose n = 2 in our experiment. The phase is scanned with a mirror mounted on a piezoelectric crystal, with a ramp signal at 3 kHz frequency. The acquisition time is chosen to be shorter than the characteristic time of the phase drift, which is around 5 ms in our experiment. 

Data are acquired at a fast data-rate, then post-processed using a freely chosen temporal mode to recover quadrature values associated to this temporal mode. This allows to explore a variety of parameters with a single set of measurements (see appendix for details). For a given temporal mode applied to the measured signal we obtain $N_{\psi} = 900$ quadrature samples ${q_{j}}$ per one scan of the phase. Changing the measured temporal mode affects both parameters $s$ and $\kappa$. In our case, their relationship can be approximated by the empirical law: $\kappa \approx 1/\sqrt{s}$. The range of squeezing reachable with our setup is [-3.4,-1.54] dB, and the range of purity is [0.46,0.68].

The highest squeezing value is obtained in the temporal mode corresponding to the eigenmode of the cavity. The reference squeezing measured in this mode with phase locking is $L_{\text{ref}} = 3.4$ dB. Implementing the Fit estimator described in Eq.~\eqref{eqn:LS} yields $L_{F} = (4.0 \pm 1.2)$ dB. This estimator is significantly noisy and biased due to insufficient statistics. Applying the moment-based estimator to the same data gives $L_{\text{MoM}} = (3.3 \pm 0.3)$ dB, which is close to the reference value, with its variance saturating the CRB.

\begin{figure}
    \includegraphics[width=1\linewidth]{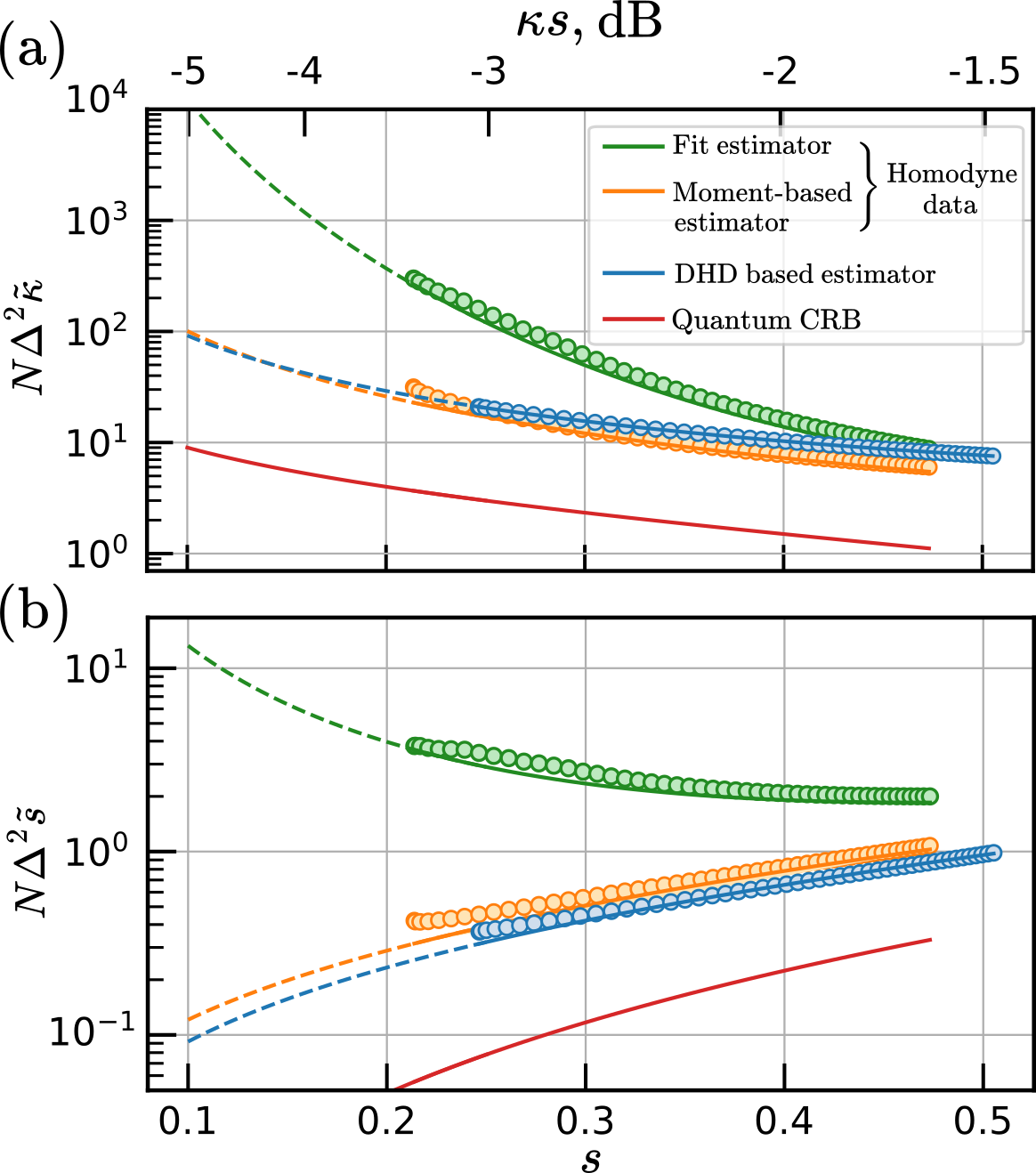}
    \caption{
    Variance of the estimator $\tilde{\kappa}$ (a) and $\tilde{s}$ (b) versus the estimated value of $s$. The markers are experimental data where the width of the dots corresponds to the error bars, the solid lines are theoretical bounds (with dashed extrapolation for smaller values of $s$).}
    \label{fig:results}
\end{figure}

The performance of the different estimators as a function of the  parameter $s$ is shown in Fig.~\ref{fig:results}. Experimental estimations are close to saturating the corresponding theoretical bounds, with some discrepancy at low $s$ values (i.e. high squeezing). The quantum Cramér–Rao bound is presented for reference; however, it cannot be achieved for all parameters simultaneously \cite{liu_quantum_2019}. The plots show that estimators obtained using the method of moments outperform those based on the fit method, with the difference becoming particularly significant at low $s$.  From experimental data, we observe up to an order of magnitude improvement, with the difference expected to increase to several orders of magnitude in the high-squeezing regime (dashed lines in the figure).

One can see that, in our case, double-homodyne detection provides more information about the parameter $s$ compared to single homodyne detection but less information about the parameter $\kappa$. However, this behavior is not universal; single homodyne detection becomes the preferred measurement for estimating all parameters in the case of highly pure states (see the appendix). Importantly, the difference between the precision of double homodyne and homodyne moment-based estimators is very small compared to the difference with the fit estimator.

\begin{figure}
    \centering
    \includegraphics[width=0.8\linewidth]{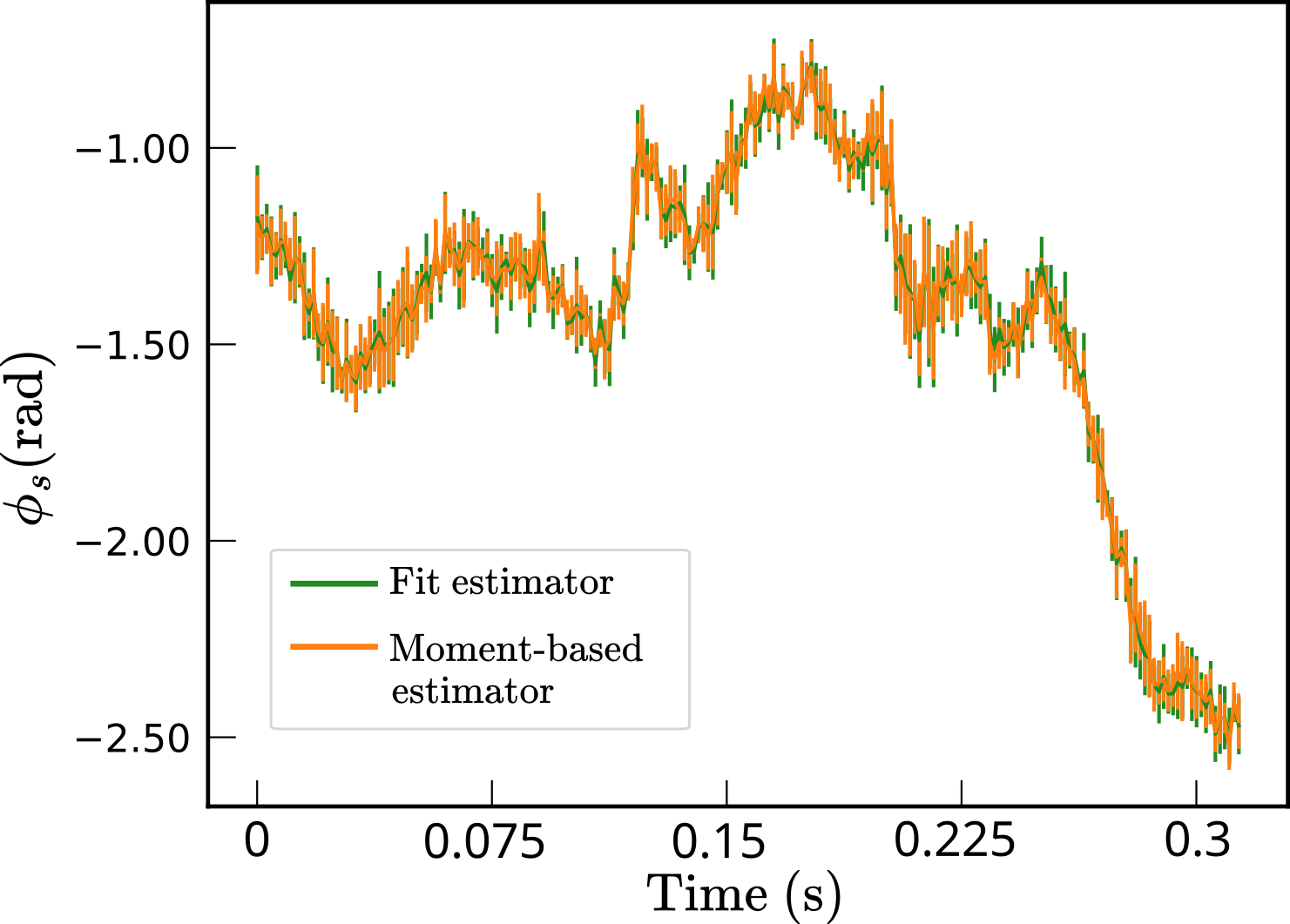}
    \caption{Estimation of the angle of squeezing $\phi_{s}$ over time with the single homodyne detection.}
    \label{fig:angle}
\end{figure}

The last estimated parameter is $\phi_{s}$, the angle of the squeezed quadrature, as shown in Fig.~\ref{fig:angle}. Oscilloscope traces were acquired during 0.3 seconds, and $\phi_{s}$ was evaluated along this measurement. The resulting data shows directly the phase noise of our experiment. This could allow to lock the phase between any quadrature of our state and the local oscillator. Contrary to typical quantum optics experiments, this would be done without using a classical seed beam. This measurement allows to measure the characteristic phase noise time of the experiment, which is around 5 ms. This confirms that the piezoelectric ramp is faster than the phase noise time, which is necessary for the method to work. 

\paragraph{Conclusion} --- In summary, we demonstrated an application of multi-parameter moment-based estimation to single-mode squeezed states. It allows to estimate the amount of squeezing, the purity, and the angle of the squeezed quadrature simultaneously and efficiently. This method outperforms the fitting approach, often used for the same task. While the range of squeezing experimentally reached in this research is limited, we show that the gain in precision grows with the squeezing level which highlights the relevance of this method. The experimentally observed precision of the proposed estimator closely approaches the CRB, indicating good agreement with the theory. The analytical form of the estimator makes it computationally efficient, potentially enabling real-time parameter tracking and drift compensation (particularly phase drifts) without requiring additional measurements or switching to a coherent seed.

From a broader perspective, we highlight that this method can be generalized for the characterization of multimode states. This problem becomes particularly challenging as the number of modes increases, making optimization strategies such as maximum likelihood nearly impossible to implement due to the high numerical complexity. 

\paragraph{Acknowledgments}--- The authors would like to thank Giacomo Sorelli for the fruitful discussions, as well as Yann Bouchereau, Leonardo Rinc\'on and Ganaël Roeland for the experimental work. This work was supported by QuantERA II project SPARQL that has received funding from the European Union’s Horizon 2020 research and innovation programme under Grant Agreement No 101017733.

\paragraph{Data availability}--- The data presented in this study are available on request.


\bibliography{lit}

\newpage

\appendix
\begin{center}    
\textit{Least squares fit estimator}
\end{center}
\vspace*{-3pt}
To find the analytical LS-estimator one can estimate the first two non-zero Fourier components of the squared quadratures as follows:
\begin{equation}
\label{eqn:c0,c2}
    \tilde C_0 = \frac{1}{N_\psi} \sum_{j=1}^{N_\psi} q^2_j,~~~ \tilde C_2 = \frac{1}{N_\psi} \sum_{j=1}^{N_\psi} q^2_j~ e^{-2 i \psi_j}.
\end{equation}
From these, the parameters $\vec \theta$ can be estimated as:
\begin{multline} 
\label{eqn:c_to_theta}
     \tilde s^F = \sqrt{ 
     \frac{\tilde C_0-2|\tilde C_2|}{\tilde C_0+2|\tilde C_2|}
     },~~~ \tilde \kappa^F = \sqrt{\frac{\tilde C_0-2|\tilde C_2|}{(\tilde C_0+2|\tilde C_2|)^{-1}}
     },\\
     \tilde \phi_s^F = -\frac{1}{2} \operatorname{Arg} \tilde C_2.
\end{multline}
It is important to note that, due to statistical noise, the estimation of the variance of the squeezed quadrature $\tilde \kappa^F \tilde s^F=\tilde C_0-2|\tilde C_2|$ can sometimes yield negative values. This indicates that the estimator fails to provide physically valid parameter values, a situation observed more frequently at higher levels of squeezing. 

To observe the variance of the estimators $\vec{\tilde \theta}^F$ we introduce variables $V_j=q_j^2$, which have expected values $\langle V_j \rangle =V(\psi_j,\vec \theta)$ and the diagonal covariance matrix $\operatorname{cov}(V_j,V_k)=\Delta^2 V_j = 2 V^2(\psi_j,\vec \theta)$, with function $ V(\psi,\vec \theta)$ defined in Eq.\eqref{eqn:V}. Then, using the error propagation principle
\vspace*{-7pt}
\begin{equation}
    \operatorname{cov} \left(\tilde \theta^F_\alpha,\tilde \theta^F_\beta \right) = \sum_{j,k=1}^{N_\psi} \left( \frac{\partial \tilde \theta^F_\alpha}{\partial V_j} \frac{\partial \tilde \theta^F_\beta}{\partial V_k} \right) \Bigg|_{V_l=\langle V_l \rangle }  \operatorname{cov}(V_j,V_k).
\end{equation}
To find this sum we precalculate the mean values of the coefficients $\tilde C_0$ and $\tilde C_2$ and their derivatives in the continuous limit $N_\psi \rightarrow \infty $:
\begin{equation}
    \langle \tilde C_0 \rangle = \frac{1}{N_\psi}\sum_{j=1}^{N_\psi} V(\psi_j,\vec \theta) \rightarrow \frac{1}{n \pi} \int_0 ^{n \pi} V(\psi, \vec \theta) d \psi 
\end{equation}
which gives 
\begin{equation}
\langle \tilde C_0 \rangle = \kappa \frac{1+s^2}{2s}, \langle  \operatorname{Re} \tilde C_2 \rangle = -\kappa \frac{1-s^2}{4s} \cos 2 \psi_s,
\end{equation}
\begin{equation}
\left \langle \frac{\partial \tilde C_0}{\partial V_j} \right \rangle = \frac{1}{N_\psi}, 
\left \langle \frac{\partial \operatorname{Re} \tilde C_0}{\partial V_j} \right \rangle = \frac{\cos 2 \psi_j}{N_\psi}.
\end{equation}
Imaginary part $\langle  \operatorname{Im} \tilde C_2 \rangle$ and its derivatives can be found analogously to $\langle  \operatorname{Re} \tilde C_2 \rangle$ by replacing $\cos$ to $-\sin$.

\begin{center}    
\textit{The method of moments}
\end{center}
\vspace{-3pt}
The performance of the moment-based estimator $\vec{\tilde{ \theta}}$ based on the observables $\vec{\hat X}$ can be assessed with help of the moment matrix \cite{gessner_multiparameter_2020}:
\vspace{-6pt}
\begin{equation} \label{eqn:M_matrix}
   { M_{\alpha \beta} (\vec \theta, \vec {\hat X}) = \frac{    \partial \langle \vec{\hat X}^T \rangle_{\vec\theta}}{\partial \theta_\alpha} \mathbf{\Gamma}^{-1} \frac{    \partial \langle \vec{\hat X} \rangle_{\vec\theta}}{\partial \theta_\beta},}
\end{equation}
where $\mathbf{\Gamma}$ stands for the measurement covariance matrix.  The covariance matrix of the estimator is given by
\vspace{-8pt}
\begin{equation}
       \operatorname{cov} \vec{\tilde \theta}= \boldsymbol{M}^{-1}.
\end{equation}

We build the estimators based on the second moments $\hat X_j = \hat q_j^2$ of the quadratures. Since $\hat q_j$ are independent and distributed as $\mathcal{N} (0, V_j)$, the covariance matrix of the moments $\vec{\hat X}$ is diagonal $\Gamma_{jk}=\delta_{jk}~ 2 V_j^2$ and the moment matrix \eqref{eqn:M_matrix} coincide with the FI \eqref{eqn:FI}.

The moment-based estimator for the Gaussian state characterization with the homodyne measurement can be expressed through the linear combinations with weights $c_\alpha (\psi_j, \vec \theta_0)$ defined in Eq.\eqref{eqn:c_formula}
\vspace{-6pt}
\begin{equation}
\label{eqn:Y_exp}
    y_\alpha (\vec \theta_0) 
    = \dfrac{1}{N_\psi} \sum_{j=1}^{N_\psi} c_\alpha (\psi_j, \vec \theta_0)~x_j,
\end{equation}
 as
 \vspace{-12pt}
 \begin{equation}
 \label{eqn:estimators}
    \begin{pmatrix}
        \tilde s \\ \tilde \kappa \\ \tilde \phi_s
    \end{pmatrix} = 
    \begin{pmatrix}
       \sqrt{\left|\dfrac{y_1(\vec \theta_0) s_0 (1+s_0) + y_2(\vec \theta_0) \kappa_0}{y_1(\vec \theta_0) (1+s_0) - y_2(\vec \theta_0) \kappa_0 } \right|}\\[16pt]
       2\kappa_0  \sqrt{\left|\dfrac{y_1(\vec \theta_0) s_0 (1+s_0) + y_2(\vec \theta_0) \kappa_0}{(y_1(\vec \theta_0) (1+s_0) - y_2(\vec \theta_0) \kappa_0)^{-1} } \right|}\\[12pt]
       {\phi_s}_0 -  \dfrac{1}{2}\dfrac{y_3(\vec \theta_0)}{y_1(\vec \theta_0) (1-s_0^2)}
    \end{pmatrix}.
\end{equation}

\vspace{-5pt}
\begin{center}    
\textit{Homodyne and double homodyne for highly pure states}
\end{center}

\begin{figure}[h]
    \centering
    \includegraphics[width=.7\linewidth]{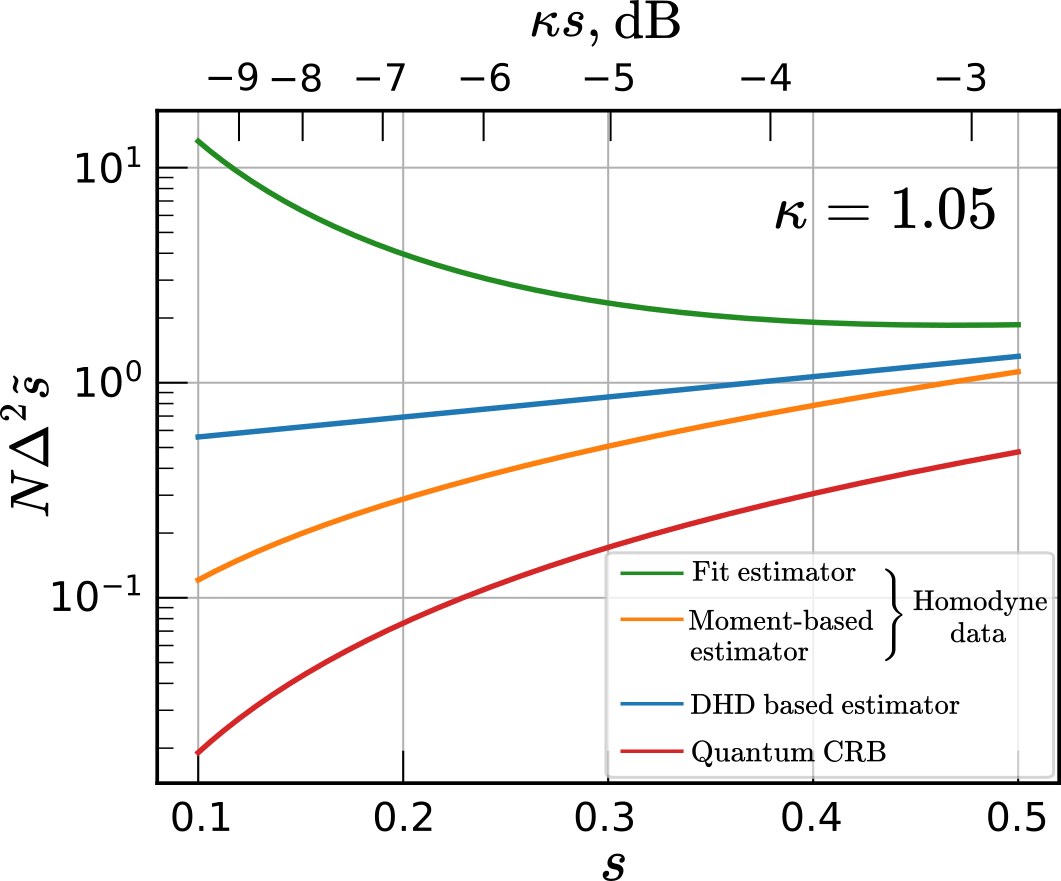}
    \caption{Theoretical bounds for different estimators in the case of highly pure states (\(95\%\)).}
    \label{fig:Bound_theory}
\end{figure}

For the squeezed light produced in our experiment, DHD outperforms homodyne detection in estimating the parameter $s$. However, for highly pure states, the situation reverses. In Fig.~\ref{fig:Bound_theory}, we theoretically compare the performance of homodyne detection \eqref{eqn:CRB} and DHD \eqref{eqn:DHD bound} for a fixed parameter $\kappa = 1.05$, corresponding to a high purity of approximately 95\%. One can see that, in this case, single homodyne detection performs noticeably better than DHD in estimating parameter $s$ (the same is true for the parameter $\kappa$).

\vspace{-4pt}
\begin{center}    
\textit{Quantum fisher information matrix}
\end{center}
\vspace{-6pt}
Quantum Fisher Information Matrix for a Gaussian state can be found as 
\vspace{-6pt}
\cite{pinel_quantum_2013}
\begin{equation} \label{eqn:QFI_matrix_Gauss}
        \boldsymbol{{\mathcal{F}}}^Q = \operatorname{diag} \left(
        \dfrac{1}{s^2}\dfrac{\kappa^2}{\kappa^2+1}, \dfrac{1}{\kappa^2-1}, \dfrac{(1-s^2)^2}{s^2} \dfrac{ \kappa^2}{\kappa^2+1}\right).
\end{equation}
Note that due to the incompatibility of optimal measurements, the quantum CRB cannot be achieved for all three parameters simultaneously \cite{liu_quantum_2019}.

\vspace{0pt}
\begin{center}    
\textit{Experimental details}
\end{center}
To generate the squeezed light we start from Ti:Sapphire laser, which produces a train of pulses at 795~nm with duration 90~fs, at repetition rate 76~MHz. The laser beam is split in two: one is used to generate the squeezed state with an OPO, and one is used for the detection scheme. 

\begin{figure}[h]
    \centering
    \includegraphics[width=0.7\linewidth]{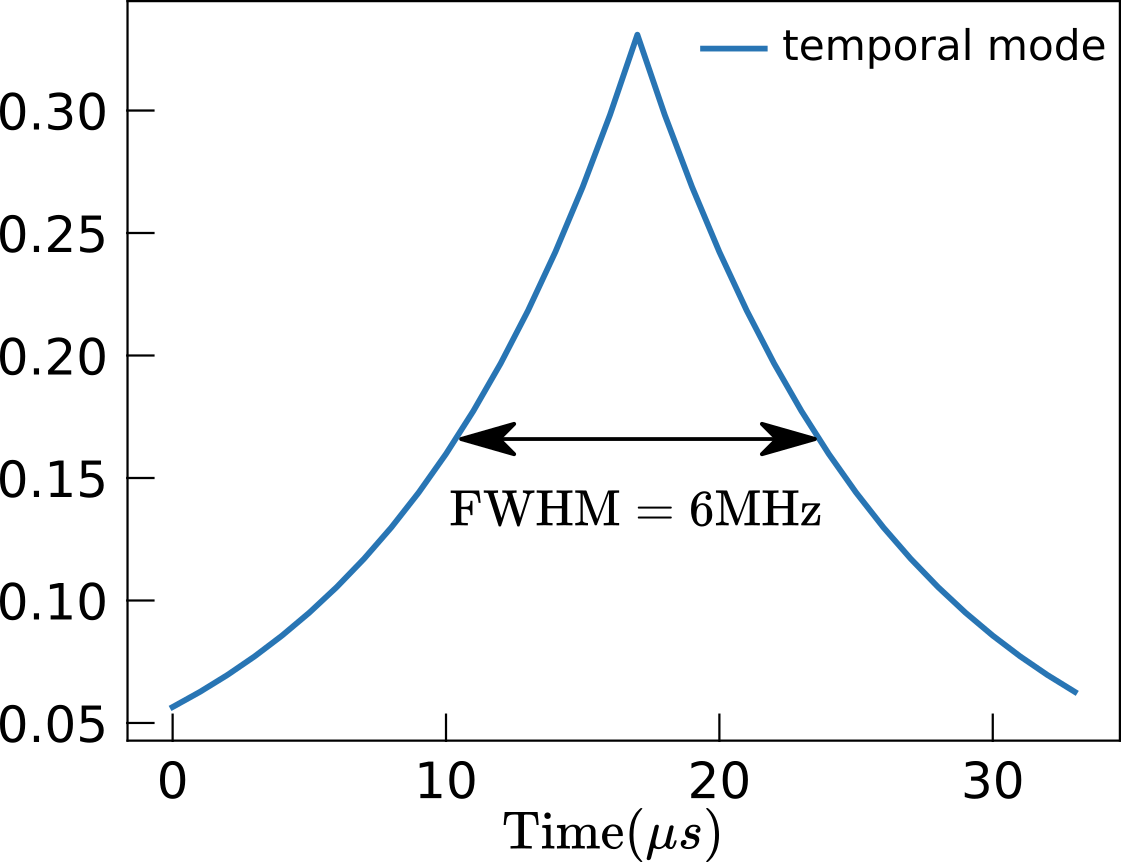}
    \caption{Temporal mode with a FWHM of 6 MHz.}
    \label{fig:temporal mode}
\end{figure}

The first beam is up-converted to a femtosecond pulse at 397.5 nm, which then enters a synchronously pumped optical parametric oscillator (SPOPO) cavity \cite{roslund2014wavelength}. This process generates a squeezed vacuum at the output. The resulting state is multimodal, and the squeezing of individual modes being limited. The length of the cavity is locked to the length of the Ti:Sapphire laser, and the transmittance of the output coupler is 50\%. The first temporal eigenmode of the cavity is a double decaying exponential, with a full width half maximum (FHWM) of about 6~MHz (see Fig. \ref{fig:temporal mode}). Another classical beam called seed is resonant with the cavity and allows for alignment of the squeezed path.

The other beam was used as the LO for both homodyne and double homodyne detection. The double homodyne detection is done in polarization: both signal and LO mix first in a PBS, the LO being circularly polarized while the signal is linearly polarized. Both arms of the PBS then encounter a homodyne detection, allowing to measure both orthogonal quadratures of the light. The polarization of the signal beam can be tuned with a half-waveplate, allowing to change between single homodyne detection and double homodyne detection easily. 

We use the interference between the LO and the seed to ensure the linearity of the piezo and uniform sampling of the quadratures over the full \(2\pi\) range. Each \(2\pi\) phase scan lasts \(500~\mathrm{\mu s}\), with an oscilloscope sampling frequency of \(100\) MHz. After applying the temporal mode to the data, we obtain \(900\) quadrature samples per scan and compute our estimators based on it. This process is repeated \(3000\) times to find the variance of the estimators.

To measure the reference of squeezing, the LO and the signal are locked through a hold and measure sequence. The seed is cut at a 3~kHz rate with a mechanical chopper. Two locks need to be active simultaneously: the first between the seed and the pump, and the second between the seed and the LO. This way, we can lock the phase between the squeezed state and the LO to only measure in the maximally squeezed quadrature. The obtained value of squeezing is $L_{ref} = 3.4$ dB.

\begin{figure}[h]
    \centering
    \includegraphics[width=0.8\linewidth]{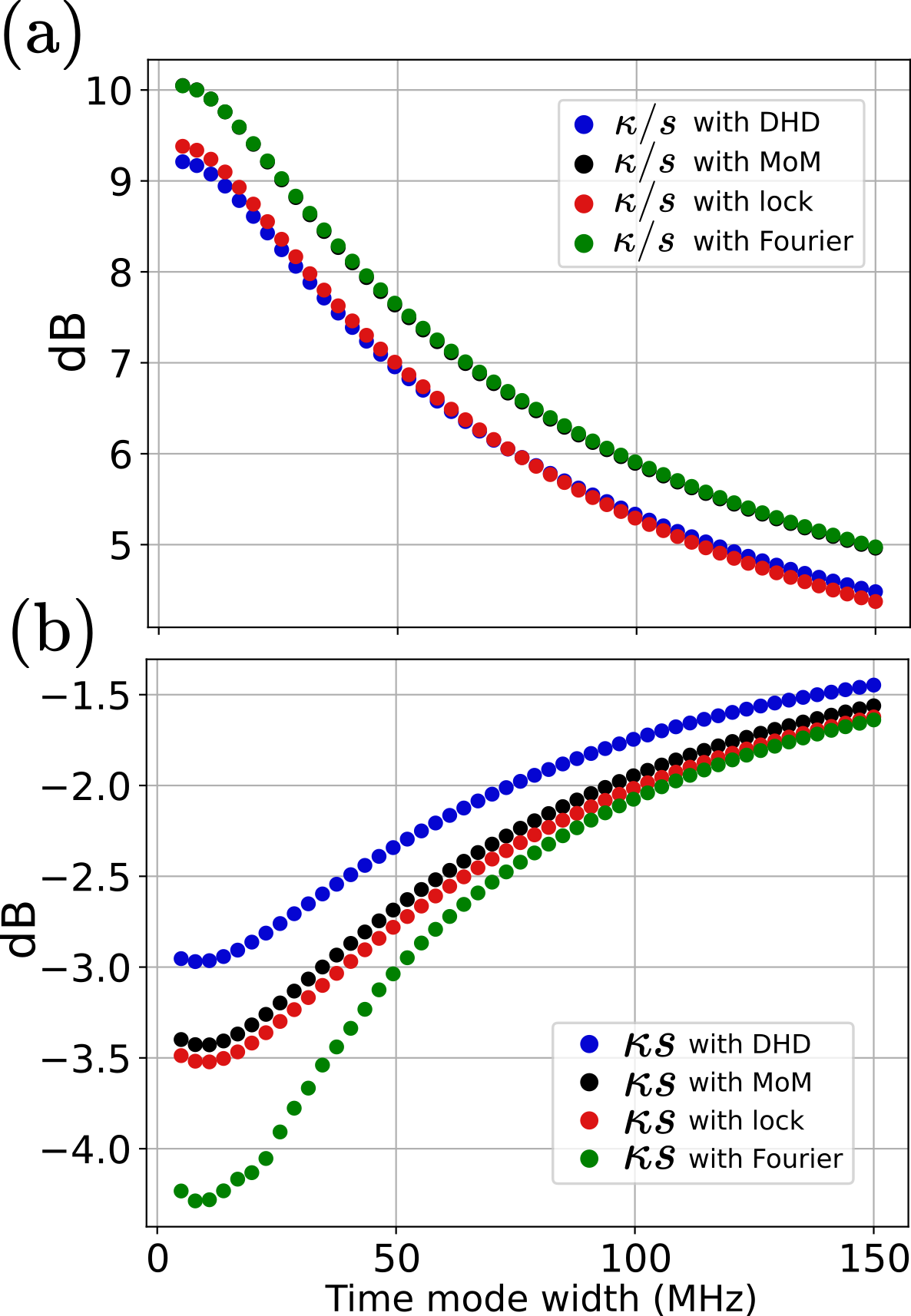}
    \caption{Evolution of the (a) antisqueezing $\kappa / s$ and (b) squeezing $\kappa s$ with the temporal mode width. The maximal squeezing is observed at the width corresponding to the temporal mode of the cavity: FWHM = 6 MHz.}
    \label{fig:sqz and antisqz}
\end{figure}

The variation of the squeezing parameter $s$ is done in post-processing, by applying the temporal mode with a different FWHM. It allows us to vary the measured mode, i.e. the value of squeezing and purity, using the same raw data from the oscilloscope. The evolution of the squeezing $\kappa s$ with the width of the temporal mode is shown in Fig. \ref{fig:sqz and antisqz}.

\end{document}